\documentclass[aps,prl,twocolumn,tightenlines,showpacs]{revtex4-2}
\usepackage[dvipdf]{graphicx}
\usepackage{color}
\usepackage{amssymb,amsmath,amsbsy}

\begin{document}
\title{Stokes force on a cylinder in the presence of fluid confinement}

\author{G.~Dolfo and J.~Vigu\'e}
 \affiliation{Laboratoire Collisions Agr\'egats R\'eactivit\'e-IRSAMC
\\Universit\'e de Toulouse-UPS and CNRS UMR 5589, Toulouse, France}
\email[]{gilles.dolfo@wanadoo.fr}
\author{D.~Lhuillier}
\affiliation{Sorbonne Université, CNRS, Institut Jean Le Rond d’Alembert, Paris, France }

\date{\today}

\begin{abstract}

In this note, we present Stokes' calculation of the force exerted by the fluid on an oscillating cylinder. While the calculation of the similar problem in the case of the sphere is treated in several textbooks, the case of the cylinder is absent from these textbooks. Because modified Bessel functions were not defined in 1851 when Stokes made this calculation, Stokes was not able to express his results in closed forms but he gave asymptotic formulas valid in the two limits  $a\ll \delta$ and $a \gg \delta$, where $a$ is the cylinder radius and $\delta$  is the viscous penetration depth. The closed form results were given by Stuart in 1963. We recall this calculation and we  compare Stokes' asymptotic formulas to these exact results. Using modified Bessel functions, it is possible to calculate the force when the fluid is confined by an external cylinder of radius $b$ sharing the same axis: we review previous publications which have treated this problem and we present an exact calculation of this force which is also developed in powers of $\delta/a$, with the expansion coefficients being functions of the ratio $\gamma= a/b$ of the cylinder radii.

\end{abstract}

\maketitle
\bigskip
 
\section{Introduction}

In 1851, Stokes \cite{Stokes1851} calculated the force exerted by the surrounding fluid on a sphere and on a cylinder in oscillating motion, in the limit of a vanishing Reynolds number. In the case of the sphere, Stokes obtained closed-form expressions of the stream function and of the force while, in the case of the cylinder, he was not able to express the stream function in closed form, because it involves modified Bessel functions which had not already been defined (following Watson \cite{WatsonBook}, these functions were defined by Basset in 1886). He was nevertheless able to give asymptotic approximations of the force in the two limits $a\ll \delta$ and $a \gg \delta$ (where $a$ is the cylinder radius and $\delta$  is the viscous penetration depth) and the exact expression of the force was given in 1963 by Stuart \cite{StuartLB63}. 

In the same paper \cite{Stokes1851}, the effect of fluid confinement was treated in the case of a sphere oscillating in a larger sphere but it was impossible to treat the similar problem of a cylinder of radius $a$ oscillating in a larger cylinder of radius $b$,  because a closed-form expression of the stream function was not available. This problem has been treated in 1976 by Chen, Wambsganss, and  Jendrzejczyk  \cite{ChenTASME76} who expressed the fluid stream function and the force exerted by the fluid on the oscillating cylinder, using modified Bessel functions. 

Because the problem of the oscillating cylinder is not treated in detail in most textbooks, we have chosen a tutorial point of view and we start our note by following Stokes' derivation as closely as possible. We then review the papers which have calculated the fluid confinement effect for the cylinder by various methods. We develop an exact calculation and, using asymptotic approximations of the modified Bessel functions, we give an expansion of the force in powers of the ratio $\delta/a$,  the expansion coefficients being functions of the ratio $\gamma= a/b$ of the cylinder radii.

\section{Navier-Stokes equations and their solution}

\subsection{Navier-Stokes equations and the stream function $\psi$}

The starting point is the Navier-Stokes equations relating the pressure $p$ and velocity $\mathbf{v}$
\begin{eqnarray}\label{S0}
\nabla p &=& \eta \Delta \mathbf{v} - \rho\left[ \frac{d\mathbf{v}}{dt} + \left( \mathbf{v}\cdot\nabla\right)  \mathbf{v} \right]  \\
\nabla\cdot \mathbf{v}&=&0. \label{S0a}
\end{eqnarray}
\noindent Here $\rho$ is the fluid density assumed to be constant and $\eta$ its viscosity. Stokes neglected the non-linear term $\left( \mathbf{v}\cdot\nabla\right)  \mathbf{v} $ because he assumed a small enough velocity: the problem then has an analytic solution. This  approximation is good if the Reynolds number $Re$ is very small, $Re\ll 1$.  

We consider a cylinder of radius $a$ oscillating in a cylinder of radius $b$. When the inner cylinder is at rest, the axes of both cylinders coincide with the $\mathbf{z}$-axis, and the oscillation is along the $\mathbf{x}$-axis, with the inner cylinder center at $x(t)= x_0 \cos \left( \omega t\right)$. The problem is restricted to 2 dimensions. Noting  $u$ and  $v$ the velocity components, eqs. (\ref{S0},\ref{S0a}) are projected on the $\mathbf{x}$- and $\mathbf{y}$-axes  
\begin{eqnarray} \label{S1}
\frac{\partial p}{\partial x} &=&  \eta\left[ \frac{\partial^2 u}{\partial x^2} + \frac{\partial^2 u}{\partial y^2}\right]  -\rho \frac{\partial u}{\partial t} \\
\frac{\partial p}{\partial y} &=&  \eta\left[ \frac{\partial^2 v}{\partial x^2} + \frac{\partial^2 v}{\partial y^2}\right]  -\rho \frac{\partial v}{\partial t} \label{S1a} \\
\frac{\partial u}{\partial x} &+& \frac{\partial v}{\partial y} =0. \label{S2}
\end{eqnarray}
\noindent Equation (\ref{S2}) proves that $d\psi$ defined by
\begin{eqnarray} \label{S3}
d\psi = u dy -vdx
\end{eqnarray} 
\noindent is an exact differential. $\psi$ is the stream function and its dimension is the product of a length by a velocity. 

\subsection{Elimination of the pressure $p$ and its expression as a function of $\psi$}

By derivation of eq. (\ref{S1}) with respect to $y$ and eq. (\ref{S1a}) with respect to $x$, one gets two expressions of $\partial^2 p/\partial x\partial y$. We write that these expressions are equal and we  thus eliminate the pressure $p$. We replace $u$  by $u= \partial \psi/\partial y$ and $v$ by  $v = - \partial \psi/\partial x$ and we get
\begin{eqnarray}\label{S4}
\left[\frac{\partial^2 }{\partial x^2} + \frac{\partial^2 }{\partial y^2}  - \frac{1}{\nu}\frac{\partial}{\partial t} \right]\left[\frac{\partial^2 }{\partial x^2} + \frac{\partial^2 }{\partial y^2} \right] \psi=0,  
\end{eqnarray}
\noindent where $\nu=\eta/\rho$ is the kinematic viscosity. The general solution of this equation is  
\begin{eqnarray}\label{S5} 
\psi = \psi_1 + \psi_2 
\end{eqnarray}
\noindent with $\psi_1$ and $\psi_2$ solutions of the following equations
\begin{eqnarray}\label{S6} 
\left[\frac{\partial^2 }{\partial x^2} + \frac{\partial^2 }{\partial y^2} \right] \psi_1 & =&0 \\
\left[\frac{\partial^2 }{\partial x^2} + \frac{\partial^2 }{\partial y^2}  - \frac{1}{\nu}\frac{\partial}{\partial t} \right]\psi_2 &=&0 \label{S6a}   
\end{eqnarray}
\noindent Using equations (\ref{S1},\ref{S1a}) and (\ref{S3}),  the pressure $p$ is expressed as a function of $\psi$ 
\begin{eqnarray}\label{S7}
d p &=& \frac{\partial p}{\partial x} dx + \frac{\partial p}{\partial y} dy \nonumber \\
&=& \eta \left[  dx \frac{\partial }{\partial y} - dy  \frac{\partial }{\partial x}\right] \left[\frac{\partial^2 }{\partial x^2} + \frac{\partial^2 }{\partial y^2}  - \frac{1}{\nu}\frac{\partial}{\partial t} \right]\psi   
\end{eqnarray} 
\noindent The term due to $\psi_2$ vanishes and we get 
\begin{eqnarray}\label{S8}
d p = \rho  \frac{\partial }{\partial t} \left(  \frac{\partial \psi_1 }{\partial x} dy -  \frac{\partial \psi_1}{\partial y} dx\right)   
\end{eqnarray} 

\subsection{Introduction of polar coordinates}

We introduce polar coordinates $r$, $\theta$ in the $\mathbf{x}$,$\mathbf{y}$ plane, $\theta = 0$ corresponding to the $\mathbf{x}$-axis, and the radial $v_r$ and tangential $v_{\theta}$ components of the velocity  
\begin{eqnarray}
x &=&r  \cos \theta \mbox{  and  }   y =r  \sin \theta \nonumber \\ 
u &=& v_r  \cos \theta -  v_{\theta} \sin \theta \mbox{  and  }
v =  v_r\sin \theta +  v_{\theta}  \cos \theta \nonumber
\end{eqnarray}
\noindent $d\psi= u  dy -v dx $  becomes $d\psi = v_r r d\theta - v_{\theta}   d r $ from which we deduce 
\begin{eqnarray}\label{S9}
v_r = \frac{1}{r}\frac{\partial \psi}{\partial \theta} \mbox{   and   } v_{\theta} = - \frac{\partial \psi}{\partial r}   
\end{eqnarray}
The 2D Laplacian in polar coordinates is
\begin{eqnarray}
\frac{\partial^2 }{\partial x^2} + \frac{\partial^2 }{\partial y^2} =\frac{\partial^2 }{\partial r^2} + \frac{1 }{r} \frac{\partial }{\partial r} + \frac{1 }{r^2} \frac{\partial^2 }{\partial \theta^2} \nonumber
\end{eqnarray}
\noindent $\psi_1$ and $\psi_2$ are solutions of
\begin{eqnarray}\label{S12} 
\left( \frac{\partial^2 }{\partial r^2} + \frac{1 }{r} \frac{\partial }{\partial r} + \frac{1 }{r^2} \frac{\partial^2 }{\partial \theta^2}\right) \psi_1   & =&0 \\
\left( \frac{\partial^2 }{\partial r^2} + \frac{1 }{r} \frac{\partial }{\partial r} + \frac{1 }{r^2} \frac{\partial^2 }{\partial \theta^2}  - \frac{1}{\nu}\frac{\partial}{\partial t}\right) \psi_2  & =&0 \label{S12a} 
\end{eqnarray}
\noindent  Equation (\ref{S8}) becomes 
\begin{eqnarray}\label{S13} 
dp=  \rho\frac{\partial }{\partial t} \left( \frac{\partial\psi_1 }{\partial r} r d\theta - \frac{1}{r}\frac{\partial \psi_1  }{\partial \theta} dr\right) 
\end{eqnarray}

\subsection{Boundary conditions }

We must write boundary conditions on the surfaces of the two cylinders of radii $a$ and $b$. Using complex notations, the velocity of the inner cylinder is  
\begin{eqnarray}\label{S13a}
\frac{dx(t)}{dt} = U \exp \left( i  \omega t\right) \mbox{   with   } U= i  \omega  x_0 
\end{eqnarray}
\noindent  Stokes assumed that the amplitude $x_0$ is very small with respect to the cylinder radius $a$ i.e. the Keulegan-Carpenter \cite{KeuleganJRNBS58} number $KC = \pi x_0/a$ verifies $KC\ll 1$. With this assumption, he wrote that the fluid velocity is equal to the cylinder velocity on the surface $r=a$
\begin{eqnarray}\label{S14}
v_r(a,\theta) &=&U \cos \theta \exp \left( i  \omega t\right)  \nonumber \\ 
v_{\theta}(a,\theta) &=& -U \sin \theta \exp \left( i  \omega t\right). 
\end{eqnarray}
\noindent The fluid velocity must vanish on the surface $r=b$ of the outer cylinder   
\begin{eqnarray}\label{S14b}
v_r (b,\theta)&=0 \nonumber \\ 
v_{\theta}(b,\theta)&=& 0. 
\end{eqnarray}
\noindent Equation (\ref{S14}) proves that $\psi_1$ and $\psi_2$ are proportional to $\sin \theta$ 
\begin{eqnarray}\label{S14a}
\psi_1 &=& U \sin \theta \exp \left( i  \omega t\right)F_1(r)\nonumber \\ 
\psi_2 &=& U \sin \theta \exp \left( i  \omega t\right)F_2(r)
\end{eqnarray}
\noindent where $F_1(r)$ and $F_2(r)$ have the dimension of a length. From eqs. (\ref{S12},\ref{S12a}), we deduce the equations verified by $F_1(r)$ and $F_2(r)$ 
\begin{eqnarray}\label{S15} 
\frac{d^2F_1}{dr^2} + \frac{1}{r}\frac{dF_1}{dr} -\frac{F_1}{r^2} &=&0  \\ 
\frac{d^2F_2}{dr^2} + \frac{1}{r}\frac{dF_2}{dr}- \frac{F_2}{r^2} -\kappa^2 F_2 &=&0, \label{S16} 
\end{eqnarray}
\noindent with $\kappa^2 = i\omega/\nu$. We define $\kappa$  by 
\begin{eqnarray}\label{S16a} 
\kappa =  \frac{1+i}{\delta}  \mbox{   with   } \delta = \sqrt{\frac{2\nu}{\omega}}.
\end{eqnarray}
\noindent $\delta$ is the viscous penetration depth. $r^n$ is an obvious solution of eq. (\ref{S15}) with $n^2=1$ i.e. $n=\pm 1$. We get
\begin{eqnarray}\label{S21} 
F_1(r)  = \frac{Aa^2}{r} + Br.
\end{eqnarray}
\noindent The introduction of the $a^2$ factor in the first term makes that $A$ and $B$ are both dimensionless. We multiply eq. (\ref{S16}) by $r^2$ and we introduce $z= \kappa r$ to get 
\begin{eqnarray}\label{S22} 
z^2\frac{d^2F_2}{dz^2} + z\frac{dF_2}{dz}- \left( z^2+ 1\right)  F_2 =0. 
\end{eqnarray}
Following Watson's book \cite{WatsonBook}, the equation 
\begin{eqnarray}\label{S22a} 
z^2\frac{d^2F_2}{dz^2} + z\frac{dF_2}{dz}- \left( z^2 + \nu^2\right)  F_2 =0 
\end{eqnarray}
\noindent has two independent solutions which are the modified Bessel functions $I_{\nu}(z)$ and $K_{\nu}(z)$ so that  
\begin{eqnarray}\label{S23} 
 F_2(r)  = CaI_1(\kappa r)+ Da K_1(\kappa r).  
\end{eqnarray}
\noindent  The $a$ factor has been introduced so that $C$ and $D$ are also dimensionless. With these results, the stream function $\psi$ is given by
\begin{eqnarray}\label{S23a} 
\psi =  \left[  \frac{Aa^2}{r} + Br + CaI_1(\kappa r)+ Da K_1(\kappa r)\right]   U \sin \theta \exp \left( i  \omega t\right). \nonumber \\
\end{eqnarray}
\noindent  Although the modified Bessel functions were not defined in 1851, Stokes was able to calculate many properties of the solutions of eq. (\ref{S16}). These results are recalled below. 

The unknown constants $A,B,C,D$ are fixed by the boundary conditions in $r=a$ and in $r=b$. Equation (\ref{S9}) gives $v_r$ and $v_{\theta}$ as a function of $F_1(r)$ and $F_2(r)$ and we then use eqs. (\ref{S14},\ref{S14b}) to get
\begin{eqnarray}\label{T6} 
F_1(a) + F_2(a) &=& a \nonumber \\ 
F_1^{'}(a) + F_2^{'}(a)&=&1 \nonumber \\ 
F_1(b) + F_2(b) &=& 0 \nonumber \\ 
F_1^{'}(b) + F_2^{'}(b)&=& 0, 
\end{eqnarray}
\noindent or in explicit form 
\begin{eqnarray}\label{T6b} 
A + B + C I_1(\alpha) + D K_1(\alpha)  &=& 1  \nonumber \\ 
-A + B + C \alpha I_1^{'}(\alpha) + D\alpha K_1^{'}(\alpha)  &=& 1  \nonumber\\
A\frac{a}{b} + B \frac{b}{a} + C  I_1(\beta) + D  K_1(\beta)  &=& 0  \nonumber \\ 
-A \frac{a^2}{b^2} + B + C \alpha I_1^{'}(\beta) + D\alpha K_1^{'}(\beta)  &=& 0  
\end{eqnarray} 
\noindent with $\alpha=\kappa a$ and $\beta=\kappa b$. For these notations and the use of dimensionless unknowns $A,B,C,D$, we have followed the paper of Chen \textit{et al.} \cite{ChenTASME76}. Before solving this system, we give the expression of the force as a function of the stream function. 

\section{Expression of the force}

By symmetry, the only non-vanishing component of the force is along the $\mathbf{x}$-axis and it is proportional to the length $l$ of the cylinder. It is given  by
\begin{eqnarray}\label{S27} 
\frac{dF_x}{dl} &=& \int_0^{2\pi} \sigma\cdot\mathbf{n} a d\theta \nonumber \\ 
\sigma_{jk}  &=&  -p \delta_{jk} + \eta\left(\frac{\partial v_j}{\partial x_k} + \frac{\partial v_k}{\partial x_j}   \right).  
\end{eqnarray}
\noindent $\mathbf{n}$ is the vector normal to the cylinder surface oriented inward $n_r= -1$ and $n_{\theta}=0$. We need only $\sigma_{rr}$ and $\sigma_{r\theta}$
\begin{eqnarray}\label{S27a} 
\frac{dF_x}{dl} &=&\int_0^{2\pi}\left(  \sigma_{rr}\cos \theta -  \sigma_{r\theta} \sin \theta \right)a d\theta
\end{eqnarray}
\noindent The components in cylindrical coordinates of the tensor $\sigma_{jk}$ are given in chapter III of the book  Laminar Boundary Layers \cite{Rosenhead63}:
\begin{eqnarray}\label{S28} 
\sigma_{rr} &=& -p + \eta\frac{\partial v_r}{\partial r} \nonumber \\ 
\sigma_{r\theta} &=&  \eta\left[r \frac{\partial}{\partial r} \left(\frac{v_{\theta}}{r} \right) +\frac{1}{r} \frac{\partial v_r}{\partial \theta}   \right] \nonumber \\ 
&=&  \eta\left[ - \frac{v_{\theta}}{r}+  \frac{\partial v_{\theta}}{\partial r} +\frac{1}{r} \frac{\partial v_r}{\partial \theta}   \right] 
\end{eqnarray}

\subsection{Calculation of $\sigma_{rr}$}

We need the derivative $\partial v_r/\partial r$ for $r=a$
\begin{eqnarray}\label{S29} 
\frac{\partial v_r}{\partial r} &=&  \frac{1}{r} \frac{\partial^2 \psi}{\partial \theta\partial r} - \frac{1}{r^2} \frac{\partial \psi}{\partial \theta} \nonumber \\ 
\left( \frac{\partial v_r}{\partial r}\right)_a &=& \frac{1}{a}\left( \frac{\partial^2 \psi}{\partial \theta\partial r}\right)_a - \frac{1}{a^2}\left( \frac{\partial \psi}{\partial \theta}\right)_a =0
\end{eqnarray}
\noindent We prove this result thanks to eqs. (\ref{S9}) and (\ref{S14}) to get $\left(\partial \psi/\partial r\right)_a = - v_{\theta}=  U \sin \theta \exp \left( i  \omega t\right)$. We then derive with respect to $\theta$ to get  $\left( \partial^2 \psi/\partial \theta\partial r\right)_a =U  \cos \theta \exp \left( i  \omega t\right)$. The radial velocity $v_r(a,\theta)$  is  related to   $\left( \partial \psi/\partial \theta\right)_a$ and we get $\left( \partial \psi/\partial \theta\right)_a= aU   \cos \theta \exp \left( i  \omega t\right)$ so that $\partial v_r/\partial r =0$. As a consequence  $\sigma_{rr} = -p$.

\subsection{Calculation of $\sigma_{r\theta}$}
\begin{eqnarray}\label{S29b} 
\sigma_{r\theta} = \eta\left[ - \frac{v_{\theta}}{r}+  \frac{\partial v_{\theta}}{\partial r} +\frac{1}{r} \frac{\partial v_r}{\partial \theta}   \right]_a 
\end{eqnarray}
\noindent $ \left( \partial v_{\theta}/\partial r\right)_a $ is given by $ \left( \partial v_{\theta}/\partial r\right)_a = -  \left( \partial^2 \psi/\partial r^2 \right)_a$.  To calculate $\left( \partial^2 \psi/\partial r^2 \right)_a$, we use the equations verified by $\psi$, $\psi_1$, $\psi_2$ 
\begin{eqnarray}\label{S32} 
-\left( \frac{\partial^2 \psi}{\partial r^2} \right)_a &=& \frac{1 }{a} \left( \frac{\partial \psi}{\partial r}\right)_a + \frac{1 }{a^2}\left(  \frac{\partial^2 \psi}{\partial \theta^2}\right)_a - \frac{1}{\nu}\left( \frac{\partial \psi_2}{\partial t}\right)_a  \nonumber \\
\end{eqnarray}
\noindent The sum of the two first terms of the r.h.s. of equation (\ref{S32}) vanishes  because $\left( \partial \psi/\partial r\right)_a = U \sin \theta \exp \left( i  \omega t\right)$ and   $\left(  \partial^2 \psi/\partial \theta^2\right)_a  =- aU\sin \theta \exp \left( i  \omega t\right)$ and we get
\begin{eqnarray}\label{S32a} 
-\left( \frac{\partial^2 \psi}{\partial r^2} \right)_a = - \frac{1}{\nu}\left( \frac{\partial \psi_2}{\partial t}\right)_a 
\end{eqnarray}
\noindent We then calculate $\left( \partial v_r/r\partial \theta\right)_a$ 
\begin{eqnarray}\label{S33} 
\left( \frac{\partial v_r}{r\partial \theta}\right)_a &=&  \frac{1}{a^2}\left( \frac{\partial^2 \psi}{\partial \theta^2}\right)_a= -\frac{U}{a}  \sin \theta \exp \left( i  \omega t\right) = \left( \frac{v_{\theta}}{r}\right)_a \nonumber \\
\end{eqnarray}
\noindent so that this term cancels the term in $-\left( v_{\theta}/r\right)_a $ in eq. (\ref{S29b}) and we get
\begin{eqnarray}\label{S30a} 
\sigma_{r\theta} = \eta \frac{\partial v_{\theta}}{\partial r}  = -\frac{\eta}{\nu}\left( \frac{\partial \psi_2}{\partial t}\right)_a = -\rho\left( \frac{\partial \psi_2}{\partial t}\right)_a
\end{eqnarray}

\subsection{Calculation of the force}

With these results, the force per unit length is given by
\begin{eqnarray}\label{S34} 
\frac{d F_x}{dl} =a \int_0^{2\pi}\left[ - p_a \cos \theta+ \rho \left( \frac{\partial \psi_2}{\partial t}\right)_a \sin \theta\right]  d\theta
\end{eqnarray}
\noindent Rather than calculating the  pressure $p_a$, Stokes integrates by parts 
\begin{eqnarray}\label{S35} 
 \int_0^{2\pi}p_a \cos \theta d\theta = p_a \sin \theta|_0^{2\pi} - \int_0^{2\pi}\frac{d p_a} {d\theta} \sin \theta d\theta 
\end{eqnarray}
\noindent The integrated term obviously vanishes because $\sin \theta$ vanishes at the bounds. We deduce $d p_a/d\theta$ from eq. (\ref{S13})
\begin{eqnarray}\label{S36} 
 \frac{d p_a} {d\theta} =  \rho a \frac{\partial }{\partial t} \left( \frac{\partial\psi_1 }{\partial r} \right)_a 
\end{eqnarray} 
\noindent We take $\partial/\partial t$ out of the integral to get 
\begin{eqnarray}\label{S37} 
\frac{dF_x}{dl} = \rho a \frac{\partial}{\partial t}\int_0^{2\pi}\left[a \left( \frac{\partial \psi_1}{\partial r}\right)_a + \left(\psi_2\right)_a\right]  \sin \theta d\theta
\end{eqnarray}
\noindent The values of $\partial\psi_1/\partial r$ and $\psi_2$ at $r=a$ are given by
\begin{eqnarray}\label{S38} 
\left( \frac{\partial \psi_1}{ \partial r}\right)_a &=& U \sin \theta \exp \left( i  \omega t\right)\left[ -A + B\right] \nonumber \\ 
\left(\psi_2\right)_a &=& U \sin \theta \exp \left( i  \omega t\right)  \left[ CaI_1(\kappa a)+ Da K_1(\kappa a\right]. \nonumber \\
\end{eqnarray}
\noindent From these results, we get the force per unit length
\begin{eqnarray}\label{S39} 
\frac{dF_x}{dl} &=& \pi a^2 \rho  U \frac{\partial  \exp \left( i  \omega t\right)}{\partial t}\left[ -A + B +  CI_1(\kappa a)+ D K_1(\kappa a) \right]  \nonumber \\
 &=&  i\omega \pi a^2 \rho  U   \exp \left( i  \omega t \right)\left[ 1-2A \right]
\end{eqnarray}
\noindent where we have used the first of the equations (\ref{T6b}) to simplify the result.  

\section{The force in the absence of confinement}

We first consider the case with $b\rightarrow\infty$ and we recall the results obtained by Stokes and by Stuart. 

\subsection{Stokes' results for the force } 

Stokes expressed the force per unit length of the cylinder in the form
\begin{eqnarray}\label{T1}
\frac{dF_x}{dl} &=& -2\pi \eta \left[  \left(\frac{a}{\delta}\right)^2 k^{'} \frac{d x}{dt}  + \frac{1}{ \omega}\left(\frac{a}{\delta}\right)^2 k \frac{d^2 x}{dt^2} \right] \\
&=& -\pi a^2 \rho \left[  \omega k^{'} \frac{d x}{dt}  +  k \frac{d^2 x}{dt^2} \right].
\end{eqnarray}
\noindent The term proportional to the velocity  $dx/dt$ is a friction term and the term proportional to the acceleration is the added mass term, which describes the inertia of the fluid following the cylinder in its motion. The identification of eq. (\ref{S39}) with eq. (\ref{T1}) relates the quantities $k$ and $k^{'}$ to the solution of the system of equations (\ref{T6b}) by
\begin{eqnarray}\label{T1b}
k - i k^{'} = 2A-1.
\end{eqnarray}
\noindent  The added mass per unit length $dm/dl$ is  given by
\begin{eqnarray}\label{T1a}
\frac{dm}{dl} &=&  \pi a^2 \rho k.
\end{eqnarray}
\noindent Stokes calculated asymptotic expansions of $k$ and  $k^{'}$  in the two limits $a\ll\delta$ and $a\gg\delta$. In the low-frequency case, $a\ll\delta$,  the quantities $\left( k -1\right)$ and $k^{'}$ diverge while these quantities multiplied by $\left( a/\delta\right) ^2$ tend toward $0$. We reproduce here the asymptotic behaviors of these last quantities
\begin{eqnarray}\label{T2a} 
\left( \frac{a}{\delta}\right)^2 \left( k -1\right)  &\approx& \frac{\pi/2}{L^2(a/\delta) + (\pi^2/4)} \\
\left(\frac{a}{\delta}\right)^2 k^{'} &\approx &  -\frac{2L(a/\delta)}{L^2(a/\delta) + (\pi^2/4)} \label{T2b} \\
\mbox{with  } L(a/\delta)&=&  -  \frac{\ln(2)}{2}+ \gamma_E +\ln\left(\frac{a}{\delta}\right), \label{T2c} 
\end{eqnarray}
 \noindent  where $\gamma_E$ is the Euler constant, $\gamma_E\approx 0.577$. In the high-frequency case, $a\gg\delta$, we reproduce Stokes's expansions of $(a/\delta)^2 k$  and  $(a/\delta)^2 k^{'}$ limited to the 3 dominant terms  (there is no constant term in  $(a/\delta)^2 k$)
\begin{eqnarray}\label{T2d} 
\left(\frac{a}{\delta}\right)^2 k  &\approx &  \left(\frac{a}{\delta}\right)^2 + 2 \frac{a}{\delta} + \frac{\delta}{8a},  \\
\left(\frac{a}{\delta}\right)^2 k^{'} &\approx &   2 \frac{a}{\delta} +1  - \frac{\delta}{8a}. \label{T2e}
\end{eqnarray}
\noindent From these results, one easily deduces the asymptotic behaviors of $k$ and  $k^{'}$ 
\begin{eqnarray}\label{T2g} 
 k  &\approx & 1 + 2 \left( \frac{\delta}{a}\right)  +\frac{1}{8}\left( \frac{\delta}{a}\right)^3,  \\
 k^{'} &\approx &   2 \left( \frac{\delta}{a}\right)  +\left( \frac{\delta}{a}\right)^2  - \frac{1}{8}\left( \frac{\delta}{a}\right)^3. \label{T2h} 
\end{eqnarray}
\noindent The added mass $dm/dl$ is then given by
\begin{eqnarray}\label{T2f}
\frac{dm}{dl} &\approx &  \pi  a^2 \rho \left[ 1 + 2\left( \frac{\delta}{a}\right)  +\frac{1}{8}\left( \frac{\delta}{a}\right)^3\right].
\end{eqnarray}
\noindent $\pi  a^2 \rho$ is the mass of displaced fluid per unit length of the cylinder and the following terms represent the contribution of the boundary layer of thickness $\delta$. In the sphere case, the expression of the added mass is fully similar \cite{Landau59} but the main term is equal only to half of the mass of the displaced fluid.

\subsection{Stuart's results}

Stuart \cite{StuartLB63} has made the calculation using modified Bessel functions. The function $I_1\left(\kappa r \right)$ diverges when  $r\rightarrow\infty$  (see ref. \cite{WatsonBook}) and this divergence, which is exponential, cannot be compensated by the divergence of $r$. This proves that $B=C=0$ and the system (\ref{T6b}) is simplified with two unknowns $A$ and $D$ 
\begin{eqnarray}\label{S24} 
A + D K_1(\alpha) &=& 1  \nonumber \\ 
-A +D \alpha K_1^{'}(\alpha)&=&1.
\end{eqnarray}
\noindent and we get 
\begin{eqnarray}\label{S24a} 
D &=& \frac{2}{K_1(\alpha) + \alpha K_1^{'}(\alpha) }  \nonumber \\ 
A &=& 1 - \frac{2 K_1(\alpha)}{K_1(\alpha) + \alpha K_1^{'}(\alpha) } = 1 + \frac{2 K_1(\alpha)}{ \alpha K_0(\alpha) }.
\end{eqnarray}
\noindent $A$ has been simplified thanks to the equality  $K_1(\alpha) + \alpha K_1^{'}(\alpha) = -\alpha K_0(\alpha)$ (see Watson's book \cite{WatsonBook}). This simplification was introduced by Hussey and  Vujacic  \cite{HusseyPF67}. We thus get
\begin{eqnarray}\label{S40b} 
k - i k^{'} = 2A-1=\left[1 +  \frac{4K_1(\alpha)}{ \alpha K_0(\alpha)}\right]  
\end{eqnarray}
\noindent We have compared numerically the expansions given by Stokes to the exact results given by eq. (\ref{S40b}). If $a/\delta <0.1$, the approximate results given by eqs. (\ref{T2a}, \ref{T2b}) are accurate  with a relative error smaller than $7$\%  for $k$ and $1$\% for $k^{'}$. If $a/\delta >1$, the approximate results given by eqs. (\ref{T2g}, \ref{T2h}) are also accurate  with a relative error smaller than $7$\%  for $k$ and $1$\% for $k^{'}$.

\section{The force in the general case}

If $b$ is finite, we need to solve the system of equations (\ref{T6b}). The derivatives of $I_1$ and $K_1$ can be replaced by modified Bessel functions either of orders $0$ and $1$ or of orders $1$ and $2$. We present these two calculations. The first one was done by Chen, Wambsganss, and  Jendrzejczyk  \cite{ChenTASME76} and, using Mathematica \cite{Wolfram}, we have done the second one which gives slightly simpler expressions.

\subsection{Calculation of  Chen, Wambsganss, and  Jendrzejczyk }

To solve the system of equations  (\ref{T6b}), Chen \textit{et al.} \cite{ChenTASME76} have replaced the derivatives $I_1^{'}$ and $K_1^{'}$ using the relations \cite{WatsonBook}
\begin{eqnarray}\label{T6e} 
z I_1^{'}(z) &=& zI_0(z)- I_1(z)  \nonumber \\ 
z K_1^{'}(z) &=& -zK_0(z)-K_1(z).  
\end{eqnarray} 
\noindent  Their results are expressed by fractions $A = A_{num}/\Delta$, $B= B_{num}/\Delta$, ... with the same denominator $\Delta$:
\begin{eqnarray}\label{chen1} 
A_{num} &=& -\alpha^2 \left[  I_0(\alpha)  K_0(\beta) -  I_0(\beta)  K_0(\alpha)\right]  \nonumber \\&& +   2 \alpha \left[  I_1(\alpha)  K_0(\beta) +  I_0(\beta)  K_1(\alpha)\right]  \nonumber \\&&-     2 \alpha \gamma \left[  I_0(\alpha)  K_1(\beta) +  I_1(\beta)  K_0(\alpha)\right]  \nonumber \\&&+     4 \gamma \left[ I_1(\alpha)  K_1(\beta) -  I_1(\beta)  K_1(\alpha)\right]  \nonumber \\
B_{num} &=&   2 \alpha \gamma \left[  I_1(\beta)  K_0(\beta)+    I_0(\beta)  K_1(\beta)\right]  \nonumber \\&&+ \alpha^2 \gamma^2 \left[  I_0(\alpha)  K_0(\beta) -  I_0(\beta)  K_0(\alpha)\right] \nonumber \\ &&- 
   2 \alpha \gamma^2 \left[  I_1(\alpha)  K_0(\beta) +  I_0(\beta)  K_1(\alpha)\right]  \nonumber \\
C_{num}&=& -2 \alpha  K_0(\beta) -  4 \gamma  K_1(\beta) + \gamma^2 \left[ 2 \alpha  K_0(\alpha) + 4  K_1(\alpha)\right]  \nonumber \\
D_{num}&=& -2 \alpha  I_0(\beta) +   4 \gamma  I_1(\beta) + \gamma^2 \left[  2 \alpha  I_0(\alpha) - 4  I_1(\alpha)\right],  \nonumber \\
\end{eqnarray} 
\noindent with the denominator $\Delta$ given by 
\begin{eqnarray} \label{chen2} 
&&\Delta = \alpha^2 \left( 1 - \gamma^2\right)  \left[  I_0(\alpha)  K_0(\beta) -  I_0(\beta)  K_0(\alpha)\right]\nonumber \\ &&+ 2 \alpha \gamma \left[ \left[ I_0(\alpha) -  I_0(\beta)\right]   K_1(\beta) +  I_1(\beta)\left[ K_0(\alpha)  - K_0(\beta) \right] \right]  \nonumber \\ &&+ 2 \alpha \gamma^2 \left[ \left(  I_0(\beta) -  I_0(\alpha) \right)  K_1(\alpha) +  I_1(\alpha) \left(  K_0(\beta) -    K_0(\alpha)\right) \right],  \nonumber \\
\end{eqnarray}
\noindent where $\gamma=a/b=\alpha/\beta$. We have found some typographical errors in the paper of Chen \textit{et al.}:  

$\bullet$ in the numerator of $B$, there is a minus sign in front of the term $I_0(\beta)  K_1(\beta)$ whereas there should be a plus sign. We have corrected this error in eq. \ref{chen1}.

$\bullet$ all the quantities $A,B,C,D$ have a sign opposite to the one we have found when we solve the system of equations (\ref{T6b}) with the same replacement of the derivatives $I_1^{'}$ and $K_1^{'}$;

$\bullet$ their equation 9 gives $H=k - i k^{'}=-2A -1$ so that the final value of the force is exact.

\subsection{Our results}

We have used Mathematica \cite{Wolfram} to solve the system of equations (\ref{T6b}). The results involve the modified Bessel functions $I_2$ and $K_2$ which is explained by another possible replacement of $I_1^{'}(z)$ and  $K_1^{'}(z)$, namely 
\begin{eqnarray}\label{T6c} I_1^{'}(z) &=& \left(I_0(z) +I_2(z) \right)/2   \\ 
K_1^{'}(z) &=& -\left(K_0(z) +K_2(z) \right)/2, \label{T6d}
\end{eqnarray}  also given by Watson \cite{WatsonBook}.  Noting  $A = A_{num}^{'}/\Delta^{'}$, $B= B_{num}^{'}/\Delta^{'}$..., we get 
\begin{eqnarray}\label{S70} 
A_{num}^{'}  &=& \alpha\beta^2\left[ I_2(\alpha)K_2(\beta)- I_2(\beta)K_2(\alpha)\right] \nonumber \\ 
B_{num}^{'}   &=& \alpha \left[ \alpha^2 \left(I_0(\beta)K_2(\alpha)- I_2(\alpha)K_0(\beta)\right) -2 \right] \nonumber\\
C_{num}^{'}   &=& 2\left[ -\alpha^2 K_2(\alpha)+  \beta^2 K_2(\beta)\right]   \nonumber \\ 
D_{num}^{'}   &=& 2\left[ -\alpha^2 I_2(\alpha)+ \beta^2 I_2(\beta)\right]   
\end{eqnarray} \noindent  
\noindent and the denominator $\Delta^{'}$ given by
\begin{eqnarray}\label{S71} 
\Delta^{'}  &=&\alpha\left[ \left( \alpha^2-\beta^2\right) I_0(\beta)K_0(\alpha) + 2\alpha I_0(\beta)K_1(\alpha) \right. \nonumber \\
&&\left. + 2\beta I_1(\beta)K_0(\alpha) -\alpha^2 I_2(\alpha)K_0(\beta) \right. \nonumber \\
&&\left. +  \beta^2 I_0(\alpha) K_2(\beta) -4\right]  
\end{eqnarray} 
\noindent In our calculation $\gamma =a/b$ does not appear because we have replaced it by $\gamma =\alpha/\beta$. We have verified that our calculation of $H=  2A -1 $ agrees with value of $H$ given by Chen \textit{et al.}.

\section{Approximate calculations of the force on a cylinder in presence of fluid confinement}

In this section, we review a series of papers treating this subject. All these papers involve an approximation.

\subsection{Calculations neglecting the fluid viscosity}

This calculation was first done in 1844 by Stokes \cite{Stokes1844}: only the added mass is not vanishing while the friction force vanishes. The added mass $dm/dl$ and the coefficient $k$ are given by
\begin{eqnarray}\label{S41} 
\frac{dm}{dl} &=&  \pi  a^2 \rho \frac{1+ \gamma^2}{1-\gamma^2} \nonumber \\
k &=&  \frac{1+ \gamma^2}{1-\gamma^2}. 
\end{eqnarray}
\noindent  When $\gamma\rightarrow 0$, the added mass tends toward  $dm/dl =  \pi  a^2 \rho$, which is is the limit of  eq. (\ref{T2f}) when $\delta$ vanishes, while the added mass $dm/dl$ and $k$ diverge when $\gamma\rightarrow 1$. 

In 1965, Hussey and Reynolds \cite{HusseyPF65}, in order to interpret experiments in superfluid helium, have calculated the effect of a cylindrical boundary on the added mass of one or two cylinders. In the case of a single cylinder, their result agrees with Stokes' result. 

\subsection{The results of Segel \cite{SegelQAM61}}

In 1961, Segel \cite{SegelQAM61} calculated the effect of fluid confinement, using conformal mapping techniques. The two cases of a low or high frequency oscillation corresponding respectively  to $a^2/\delta^2 \ll 1$ or $ \gg 1$ were treated separately.  We report here only the results corresponding to the high frequency case $a^2/\delta^2 \gg 1$ because the results in the low-frequency case have complicated expressions. Equation 5.25 of Segel's paper gives the modification of  the coefficients $k$ and $k^{'}$ approximately given by eqs. (\ref{T2d}, \ref{T2e}) by the presence of the outer cylinder. We have expressed Segel's results with our notations 
\begin{eqnarray}\label{S51}
k &\approx &  \frac{1+ \gamma^2 }{ 1- \gamma^2}+ \frac{2}{ 1- \gamma^2}\times \left( \frac{\delta}{a}\right) \nonumber \\ 
k^{'} &\approx &  \frac{2}{1- \gamma^2} \times \left( \frac{\delta}{a}\right).  \label{S52}
\end{eqnarray}
\noindent When $\gamma \rightarrow 0$, these results converge toward the first two terms of Stokes' result for $k$  given by eq. (\ref{T2g}) and to the first term of Stokes' result for $k^{'}$ given by eq. (\ref{T2h}). In the limit of vanishing viscosity (i.e. when $\delta/a \rightarrow 0$), the coefficient $k$ agrees with the result of Stokes \cite{Stokes1844} given by eq. (\ref{S41}).

\subsection{The results of Siniavskii, Fedotovskii and Kukhtin }

In 1980, Siniavskii, Fedotovskii and Kukhtin \cite{SiniavskiiSAM80} developed an approximate calculation valid if $(b-a)\gg \delta$. They treated separately the boundary layer in which the viscosity is taken into account while the viscosity is neglected outside this layer.  When  $b\rightarrow\infty$, their calculation gives
\begin{eqnarray}\label{S61}
k &\approx &  1 + 2 \left( \frac{\delta }{a} \right) + \frac{1}{4}\left( \frac{\delta}{a}\right)^2    \\ 
k^{'} &\approx &   2 \left( \frac{\delta}{a}\right) .  \label{S62}
\end{eqnarray}
\noindent The coefficient $k$ agrees with Stokes' result given by eq. (\ref{T2f}) for the first two terms but not for the third one and the coefficient $k^{'}$ given by eq. (\ref{S62}) agrees with the leading term of Stokes' result given by eq. (\ref{T2h}). If $b$ is finite, their calculation gives the following result for the coefficient $k$
\begin{eqnarray}\label{S63a}
k &\approx &  \left( 1+ \frac{\delta}{2a}\right)^2 \frac{\left( 2b - \delta\right)^2 + \left( 2a + \delta\right)^2 }{\left( 2b - \delta\right)^2 - \left( 2a + \delta\right)^2 }  + \frac{\delta}{a}.  
\end{eqnarray}
\noindent An expansion of this result in powers of $\delta/a$ gives 
\begin{eqnarray}\label{S63b}
k  &\approx &  \frac{1+ \gamma^2}{1- \gamma^2} + \frac{2\left( 1-\gamma+ \gamma^2\right) }{\left(1+\gamma \right) \left(1-\gamma\right)^2 }\times \left( \frac{\delta}{a}\right).   
\end{eqnarray}
\noindent In the limit  $\left( \delta/a\right)\rightarrow0$, this result agrees with the well established result \cite{Stokes1844,HusseyPF65}  given by eq. (\ref{S41}). The friction force coefficient $ k^{'}$ is given by 
\begin{eqnarray}\label{S63c} 
 k^{'} &\approx &   2  \frac{1+\gamma^3}{\left( 1-\gamma^2\right)^2} \left( \frac{\delta}{a}\right).
\end{eqnarray}
\noindent We have noticed that $\left( 1+\gamma^3\right) /\left( 1-\gamma^2\right)^2 = \left( 1-\gamma+ \gamma^2\right)/\left(1+\gamma \right)$ so that the term linear in $\left( \delta/a\right)$ is the same in $k$ and in $ k^{'}$.

\subsection{Expansion of the exact results }

Chen, Wambsganss, and  Jendrzejczyk \cite{ChenTASME76} have  given a approximate form of $H $ in the limit where $\alpha = \left(1+i \right)a/\delta $ and $\beta=\left(1+i \right)b/\delta $ are both large. Their derivation is based on the asymptotic expansions of the modified Bessel functions $I_n(z)$ and $K_(z)$ given by Watson \cite{WatsonBook} :
\begin{eqnarray}\label{S80} 
I_n\left(z \right) & \sim & \exp\left( z\right)\sqrt{\frac{1 }{2\pi z}} \nonumber \\ &&\times\left(1 -\frac{4 n^2 -1}{1!8z}  + \frac{\left( 4 n^2 -1^2\right)\left( 4 n^2 -3^2\right) }{2!\left( 8z\right)^2}+ ...\right) \nonumber \\ 
K_n\left(z \right) & \sim &  \exp\left( z\right)\sqrt{\frac{\pi  }{2 z}} \nonumber \\ &&\times\left(1 +\frac{4 n^2 -1}{1!8z}  + \frac{\left( 4 n^2 -1^2\right)\left( 4 n^2 -3^2\right) }{2!\left( 8z\right)^2} +...\right) \nonumber \\
\end{eqnarray}
\noindent We have limited these expansions to the first terms of the $1/z$ series and we have neglected terms which are exponentially small if $Re\left(z \right)$ is positive and large. If we understand correctly what has been done by Chen \textit{et al.}, the series appearing in eq. (\ref{S80}) have been limited to the term equal to $1$. Their result involves terms in $\sinh\left(\beta-\alpha \right)$ and $\cosh\left(\beta-\alpha \right)$ as well as two terms in  $\alpha \gamma^{1/2}$ and $\alpha \gamma^{3/2}$ which come from the terms of the type $I_n\left(z \right)K_n\left(z \right)$ with the same $z$ ($z= \alpha$ or $z=\beta$). When $Re\left(\beta-\alpha \right)$ is positive and large, these terms are negligible with respect to the hyperbolic sine and cosine terms.  Moreover, the difference between $\sinh\left(\beta-\alpha \right)$  and $\cosh\left(\beta-\alpha \right)$ is also negligible in this case. Finally, Chen \textit{et al.} have not deduced from their calculation of $H$ the values of $k$ and $k^{'}$ as a function $\delta/a$ and $\gamma$.

\begin{figure}[h]
\begin{center}
\includegraphics[width= 8cm]{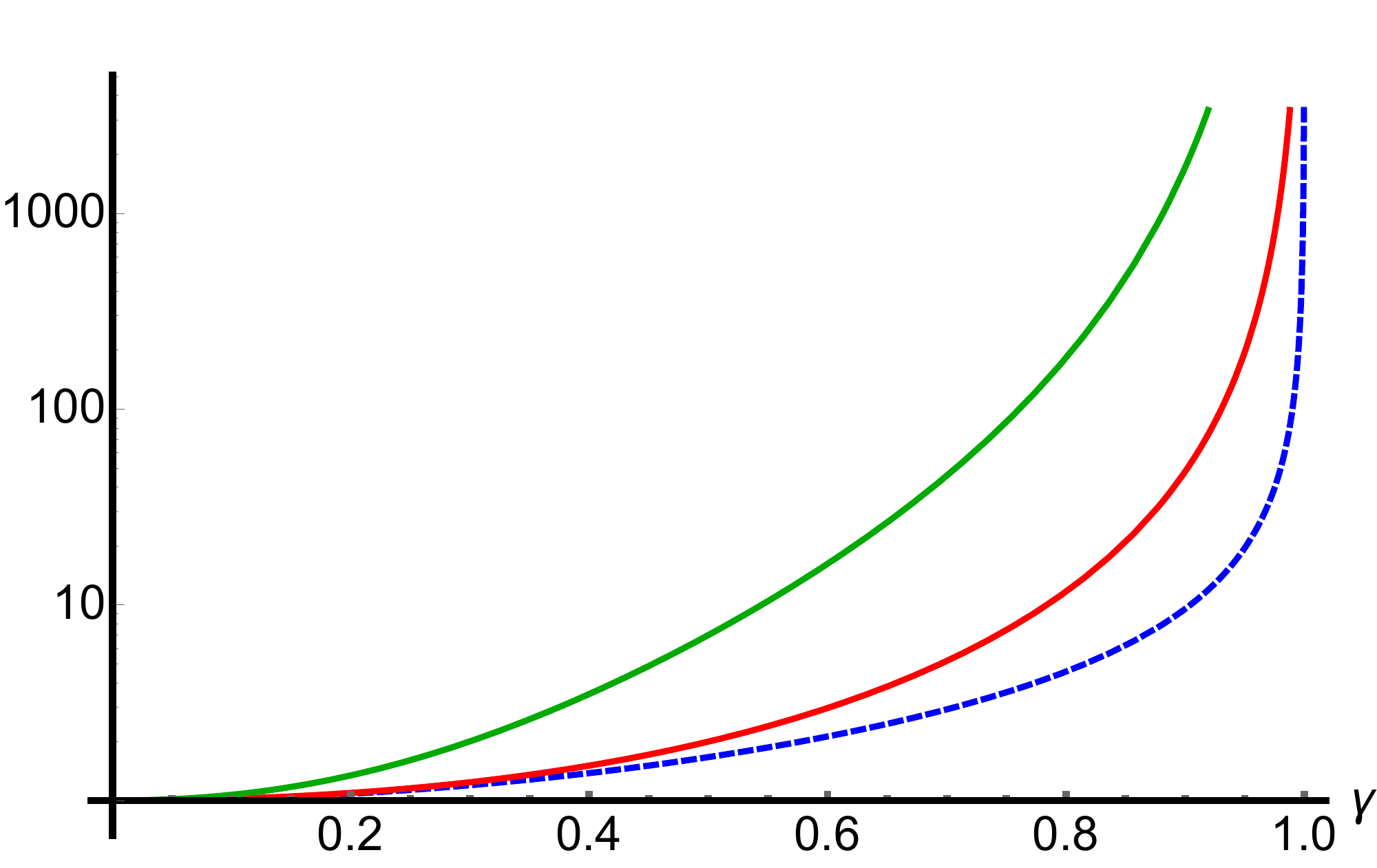}
\caption{Semi-logarithmic plot of the power expansion coefficients $F_0\left( \gamma\right) = \frac{1+ \gamma^2  }{1- \gamma^2 }$,  $F_1\left( \gamma\right) = \frac{\left(1-\gamma + \gamma^2 \right)}{\left(1+ \gamma \right)\left( 1-\gamma\right)^2} $  and $F_2\left( \gamma\right) = \frac{1 -2\gamma+ 6\gamma^2 -2\gamma^3+\gamma^4}{ \left(1+\gamma \right)\left( 1-\gamma\right)^3 }$ as a function of $\gamma$: $F_0$ is represented by the dashed (blue) curve, $F_1$ is represented by the full (red) curve and $F_2$ is represented by the full (green) curve. }\label{figure1}
\end{center}
\end{figure}

Using Mathematica \cite{Wolfram}, we have redone this calculation, including the terms up to $1/z^2$ of the series of eq. (\ref{S80}) and neglecting the negligible quantities discussed in the previous paragraph.  We have expressed $ \beta =\alpha/\gamma$ and we have expanded the results in powers of $1/\alpha =\left(1-i \right)\delta/a  $  up to the second order.  We have verified that this procedure gives the same result using the expression of $H$ obtained either by  Chen \textit{et al.} or  by ourselves. We thus get the values of $k $ and $ k^{'} $   up to the second in $\delta/a$, with the expansion coefficients being functions of $\gamma$:
\begin{eqnarray} \label{S83} 
k &=&  \frac{1+ \gamma^2  }{1- \gamma^2 } + \frac{\left(1-\gamma + \gamma^2 \right)}{\left(1+ \gamma \right)\left( 1-\gamma\right)^2} \times  \frac{2\delta}{a} +  O\left[ \left(  \frac{\delta}{a}\right)^3\right]\nonumber \\
k^{'}&=&  \frac{\left(1-\gamma + \gamma^2 \right)}{\left(1+ \gamma \right)\left( 1-\gamma\right)^2}\times  \frac{2\delta}{a} \nonumber \\&&+ \frac{1 -2\gamma+ 6\gamma^2 -2\gamma^3+\gamma^4}{ \left(1+\gamma \right)\left( 1-\gamma\right)^3 } \left( \frac{\delta}{a} \right)^2 + O\left[  \left(  \frac{\delta}{a}\right)^3\right].\nonumber \\
\end{eqnarray}
\noindent  Here are some comments:

$\bullet$ in the expansion of $k$, the term in $(\delta/a)^2$ vanishes;

$\bullet$ we have verified that the expansions of $k $ and $ k^{'} $  given by eqs. (\ref{S83}) are stable if we increase the number of terms of the  $1/z^n$ series of eq. (\ref{S80});

$\bullet$ the limits of  $k$ and $k^{'}$ when $\gamma \rightarrow 0$ are in perfect agreement with the expansions in $\left( \delta/a\right)$ given by eqs. (\ref{T2g},\ref{T2h}) corresponding to case $\gamma = 0$;

$\bullet$ the limit of $k$ when $\left( \delta/a\right)$ vanishes agrees with Stokes' result given by eq. (\ref{S41});

$\bullet$ the results of Segel \cite{SegelQAM61} are in agreement with Stokes' result for the added mass in an inviscid flow but the first order terms in $\left( \delta/a\right) $ in the expansions of $k$ and $k^{'}$ are not in agreement with the results of Siniavski \textit{et al.} which  agree with our results. However our results extend up to a second order term in $\left( \delta/a\right)$;

$\bullet$ figure \ref{figure1} presents a plot of the variations of the three non-vanishing coefficients  appearing in eqs. (\ref{S83}) as a function of $\gamma$.  They diverge when $\gamma\rightarrow 1$, the divergence of the coefficient of $\left( \delta/a\right)^n$ being due to the denominator $\left( 1-\gamma\right)^{(n+1)}$ so that the divergence is faster when $n$ increases;

$\bullet$ our procedure can give higher order terms of the expansions of  $k$ and $k^{'}$ in powers of $\left( \delta/a\right) $. However, when $\gamma$ approaches $1$, the coefficients of $\left( \delta/a\right)^n$ appear to diverge more rapidly when the order $n$ increases. As a consequence, the use of this expansion is probably limited to lower $\gamma$ values if the higher order terms of these expansions become important and it is probably better to use the exact results.

\section{Concluding remarks}

In this note, we have first reproduced Stokes' calculation of the force exerted by the surrounding fluid on a cylinder in oscillating motion. We have verified that Stokes asymptotic results are in good agreement with the exact results obtained by Stuart, using modified Bessel functions. 

In his 1851 paper, Stokes calculated the effect of confinement for a sphere oscillating inside a larger sphere but, at that time, it was not possible to make the same calculation for a cylinder oscillating inside a larger cylinder because this calculation requires the use of modified Bessel functions. This calculation was first done by Chen, Wambsganss, and  Jendrzejczyk \cite{ChenTASME76} in 1976 and, using Mathematica \cite{Wolfram}, we have reproduced their calculation and pointed out some misprints in their paper. Moreover, as Mathematica \cite{Wolfram} uses different relations between modified Bessel functions, we have obtained somewhat simpler expressions of the confinement effect. 

We report the results of the approximate calculations of the confinement effect made by Segel \cite{SegelQAM61} and by Siniavskii, Fedotovskii and Kukhtin \cite{SiniavskiiSAM80}. We have also completed the calculation of Chen, Wambsganss, and  Jendrzejczyk \cite{ChenTASME76} and we have obtained a power expansion in $\delta/a$ of the two coefficients $k$ and $k^{'}$ respectively describing the inertial part (the added mass term) and the friction part of the force. The coefficients of the power expansions of $k$ and $k^{'}$ are expressed as functions of the ratio $\gamma=a/b$. The results of Siniavskii \textit{et al.} agree with our results which also involve the next order terms and which could be extended to higher orders.

Experimental tests of the confinement effect on the added mass and on the friction force have been done by Chen \textit{et al.} \cite{ChenTASME76} and also by Siniavskii \textit{et al.} \cite{SiniavskiiSAM80}. In both cases, the experimental results have been found in good agreement with their calculations. We have also used our calculation of the confinement effect to analyze our measurements of the friction force on a cylinder oscillating inside another cylinder \cite{DolfoPRF80}: the confinement correction appears to be quite necessary for a correct interpretation of the experimental results.


\end{document}